\documentclass[12pt,draftcls,onecolumn]{IEEEtran}

\ifCLASSINFOpdf
\else
\fi

\ifCLASSOPTIONcompsoc
\usepackage[nocompress]{cite}
\else
\usepackage{cite}
\fi

\usepackage{amsmath}
\usepackage{amssymb}
\usepackage{graphicx}
\usepackage{epstopdf}
\usepackage{mathtools}
\usepackage{xcolor}

\begin{document}

\title{Rethink Delay Doppler Channels and Time-Frequency Coding}


\author{
        Xiang-Gen Xia, \IEEEmembership{Fellow}, \IEEEmembership{IEEE} 

\thanks{
  X.-G. Xia is with the Department of Electrical and Computer Engineering, University of Delaware, Newark, DE 19716, USA (e-mail: xxia@ece.udel.edu).
}

}

\date{}

\maketitle


\begin{abstract}
  In this paper, we rethink  delay Doppler channels (also called
  doubly selective channels). 
 We prove that 
 no modulation schemes (including the current active VOFDM/OTFS) can compensate
 a non-trivial Doppler spread well.
 We  then discuss some of the existing
  methods to deal with time-varying channels, in particular time-frequency
  (TF)   coding in an OFDM system. TF coding
  is equivalent to space-time coding in the math part.
  We also summarize state of the art on space-time coding that was an active
  research topic over a decade ago.
\end{abstract}



\section{No Modulation Scheme Can Compensate Non-Trivial Doppler Shifts}\label{sec1}

A doubly selective channel, i.e., it has both time spread and Doppler
spread, has re-attracted significant attention lately due to
the recent Starlink success. In fact, such a channel was studied
in the 1990's, see for example  \cite{hahm, xia0}.
A douly selective channel is also called a delay Doppler channel
\cite{sayeed}. 
A recent active 
topic is orthogonal time frequency space (OTFS) modulation \cite{otfs1, otfs4}
that has been shown identical to vector OFDM (VOFDM)  \cite{xia1, xia2, xiabook}
in \cite{otfs2, otfs3, xia3, osdm}. 
OTFS has been claimed to be able to deal with a delay Doppler channel well,
which, we think, is mis-leading and 
 no modulation scheme can well compensate a non-trivial
Doppler spread, as we shall see in details below. 

A delay Doppler channel can be described as follows \cite{hahm, xia0, sayeed}.
At time delay $\tau$, let
\begin{equation}\label{1.0}
h(\tau,t)=g(\tau) e^{-j\Omega(\tau)t}
\end{equation}
be its channel response with Doppler shift $\Omega(\tau)$ that is a function of time delay $\tau$.
It means that the path $h(\tau,t)$ of time delay $\tau$ has Doppler shift  $\Omega(\tau)$ and in general, different paths at different time delays may have different Doppler shifts.

Let $s(t)$ be a
transmitted signal. Then, the received signal $y(t)$ at time $t$  is
  \begin{eqnarray}
    y(t)& = & \int h(\tau,t)s(t-\tau)d\tau+w(t)  \nonumber\\
       & =& \int g(\tau) s(t-\tau) e^{-j\Omega(\tau)t} d\tau + w(t),\label{1.1}
  \end{eqnarray}
  where $w(t)$ is the additive noise.

 When the Doppler shift function $\Omega(\tau)$ in (\ref{1.1}) is
  a constant $\Omega$
  that does not depend on $\tau$, which is called
   trivial Doppler spread case,
  it means that all the channel responses
  at all the time delays  have the same Doppler shift $\Omega$. In this case,
  it is easy to see that
  this Doppler shift can be compensated at either transmitter or receiver
  and the compensated channel then becomes a time spread only channel.
  This argument applies to the channels on any finite time interval where
   the Doppler shift function $\Omega(\tau)$ is approximately
   constant, i.e., it can be approximated by a constant independent of
   time delay variable $\tau$. 

  Otherwise, different multipaths have different
  Doppler shifts and it is called non-trivial Doppler spread case.
  In this case, the received signal in (\ref{1.1}) becomes
  \begin{equation}\label{1.2}
    y(t) =   \int g(t-\tau) s(\tau) e^{-j\Omega(t-\tau)t} d\tau + w(t).
      \end{equation}
  For example, when the Doppler shift function $\Omega(\tau)$ is linear
  in terms of $\tau$, i.e., $\Omega(\tau)=\Omega \tau$ for some non-zero
  constant $\Omega$, then the received signal in (\ref{1.2}) is
  \begin{equation}\label{1.3}
    y(t) =   \int g(t-\tau) s(\tau) e^{j\Omega\tau t} d\tau
    e^{-j\Omega t^2}+ w(t).
  \end{equation}
  From the above signal model, the non-trivial
  Doppler spread part is also a function
  of  variable $t$,
  while the transmitted signal $s(\tau)$ only depends on $\tau$
  and no matter what $s(\tau)$ is, it has nothing to do with
  the variable $t$ in the Doppler spread, thus any transmit signal
  $s(t)$ cannot compensate the Doppler spread, no matter what modulation
  scheme is used. This conclusion holds  for both continuous 
  and discrete time signal models, and also for signal models on any time
  interval that could be short. 
  With the above result, neither VOFDM/OTFS nor GFDM \cite{GFDM, DGT, xia4}
  can compensate a non-trivial Doppler spread as also briefly mentioned in
  \cite{DDT} from a joint time-frequency analysis point of view. 

  A delay Doppler channel is a time-varying channel. Although there is
  no modulation that can compensate the non-trivial Doppler
  spread well, there were methods to deal with general time-varying channels
  to have improved performance over 
  20 years ago. The basic idea is to use a block of time slots together
  in demodulation or decoding. These methods include
  bit-interleaved coded modulation (BICM) \cite{bicm} and signal space diversity for narrow band systems \cite{ssd1,ssd2}, which can be applied along the frequency components 
   in broadband OFDM systems, 
   and time-frequency (TF) coding for broadband OFDM systems.
   Mathematically TF coding is equivalent to space-time (ST) coding
   and both of them are the two special cases of
   space-time-frequency (STF) coding. \cite{stfc} is a tutorial paper 
   about ST/SF/STF coding, where one can find the related original
   works on this topic.

   Note that for TF coding, multiple OFDM symbols
   may be coded and decoded together. If only one OFDM symbol is considered,
   information symbols in one OFDM symbol can be coded/decoded together
   by using signal space diversity along the frequency index as mentioned
   earlier. The main problem for signal space diversity and TF coding
   is the demodulation/decoding complexity, since all the information symbols
   in  one or multiple  OFDM symbols are decoded/demodulated together.

   If one OFDM symbol is demodulated together, firstly we think that
   it is againist the motivation to emply OFDM
   with simple demodulation complexty, and secondly, to deal with
   time-varying channels, one could randomly generate an $N\times N$ unitary
   matrix and apply it as a precoding before the OFDM of $N$ subcarriers.
   This may not perform as good as the signal space diversity
   technique in the frequency domain that can achieve full multipath
   diversity in theory,
   but  may perform as good as any existing linearly modified
   OFDM systems (XFDM, XYDM, XYZM), all of which may have too
   high demodulation complexities.

   Although VOFDM/OTFS cannot compensate a non-trivial Doppler
   spread, it does have improved performance for time-varying
   channels over OFDM. This is because for VOFDM/OTFS, a vector
   of information symbols are demodulated together
   \cite{xia2, xiabook, zhangxia, xia5, xia3}. For example,
   the frequency domain equalizer along a vector of symbols
   can be used. However, this performance improvement is not because
   of the particular delay Doppler channel but applies to any time-varying
   channel.

   In addition to its good performance for time-varying channels,
   VOFDM is the most general modulation in terms
   of dealing with intersymbol interference
   (ISI) channels. Assume $M$ is the vector size of VOFDM and $N$ is the
   number of subcarriers as in OFDM. For an ISI channel $H(z)$, at the
   receiver, after the cyclic prefix removal, the received
   signal becomes \cite{xia2}
   \begin{equation}\label{1.4}
     {\bf y}_k={\bf H}_k {\bf x}_k+{\bf w}_k,
   \end{equation}
   for $k=0,1,...,N-1$, and ${\bf x}_k$ and ${\bf y}_k$ are
   the $M\times 1$ information
   symbol vector and the $M\times 1$ received signal vector (after
   the component-wise $N$-point FFT of the received signal vectors),
   respectively, 
      and ${\bf w}_k$ is the additive noise vector, 
   at the $k$th frequency component. And 
   ${\bf H}_k$ is the $M\times M$
   pseudo-circulant channel matrix ${\bf H}(z)$
   evaluated at the $k$th frequency of $z=\exp(j2\pi k/N)$.
   The pseudo-circulant matrix ${\bf H}(z)$ is obtained from the
   polyphase components of the 
   original ISI channel function $H(z)$ and its more details are refered to
   \cite{xia2}. 
   The received signal model in (\ref{1.4}) was mathematically new
   in the literature for an ISI channel, which is the generalization
   of the OFDM received signal model in terms of ISI level.
   When the vector size $M=1$,
   VOFDM returns to the conventional OFDM and the received
   signal model (\ref{1.4}) is the same as the conventional one of the OFDM.
   When $M$ is not smaller
   than the ISI size and $N=1$, it returns to the conventional single
   carrier frequency domain equalizer (SF-FDE) \cite{xia5}. Thus, VOFDM
   is also a bridge between OFDM and SC-FDE. For a general vector
   size $M$ and each $k$, there are at most $M$ symbols interfering
   each other inside each vector subchannel (\ref{1.4}) but there is
   no ISI across vector subchannels, i.e., across $k$.
   Thus, for vector size $M=1,2,3,...$, VOFDM is the most general
   modudlation scheme in terms of ISI level
   that converts an ISI channel to multiple independent 
   subchannels each of which may have no ISI, $M=2$ symbols in ISI, $M=3$
   symbols in ISI, $...$, respectively.

   It is important to
   note that one can obviously  add pulses to VOFDM in transmission,
   the same as  adding pulses to OFDM in transmission. The key to
   determine whether a modulation is the same as VOFDM is to see whether
   it leads to the same receive signal model (\ref{1.4}) at the receiver
   for any fixed ISI channel, when the rectangular pulse is used. 
   
   Recall that dealing with ISI channels has been always
   the most important physical layer
   task in digital communications in the past, no matter whether in wired
   or wireless systems. In fact, this is also the most important
   in radar applications with broadband waveforms.
   Although in radar applications there is no concept of ISI,
   it is called inter-range cell interference (IRCI)
   \cite{xia_radar1, xia_radar2, xia_radar3}. The reason is simple and it  
   is because both of them use EM waves to transmit and receive. 
   Thus,  in our opinion, as the most general modulation
   to deal with an ISI channel, VOFDM/OTFS plays a more important role
   over an ISI channel than over a time-varying channel (or, in particular, 
    a delay Doppler channel).

\section{Time-Frequency Coding}\label{sec2}

TF coding is an old concept that can be done
by utlizing the signal space diversity technique \cite{ssd1,ssd2} along the
frequency index in one OFDM symbol as mentioned before
or two dimensional coding 
across multiple OFDM symbols similar to ST coding that can be thought
of a special case of space-time-frequency (STF) coding and more details
can be found in \cite{stfc}. Also, a signal space diversity design is
equivalent to a diagonal space-time block code design \cite{dstbc}.

As mentioned earlier,
the major problem for these approaches using TF/STF coding is the high
demodulation/decoding complexity. Otherwise, no modulations (XFDM, XYDM, XYZM)
can perform better than frequency domain signal space diversity 
or TF/STF coding over time-varying channels including
delay Doppler channels.
A key for these techniques is ST code design that can be briefly described
below.

An ST code is a collection of the same size matrices,  which are mapped
to bits, to transmit. One dimension of a matrix in an ST code
corresponds to time and the other dimension corresponds to
space, i.e., transmit antennas.
Without loss of generality, let us assume all the matrices
are squared, i.e., the dimensions of time and space are the same. 
If the maximum-likelihood
(ML) decoding is used at the receiver, an ST code achieves
full diversity if any difference matrix of any two distinct matrices
in the ST code has full rank. The minimum of the
absolute determinant value of all the difference matrices
of two distinct matrices in the ST code corresponds to the
coding gain (or called diversity product),
whose maximum is desired in a design of ST codes, called
optimal ST codes. Such an optimal $2\times 2$ unitary code (i.e.,
each matrix in the code is unitary) of
size $6$  is obtained in \cite{haiquan2} and a best known
$2\times2$ unitary code of size $16$ is obtained in \cite{liang}. 

The most well-known (also one of the earliest)
ST block code is the Alamouti code \cite{alamouti} for two
transmit antennas, which corresponds to
complex numbers for real information symbols and quaternions for
complex information symbols. It has been generalized to
orthogonal space-time block codes (OTSBC) for a general number of
transmit antennas \cite{ostbc}. The orthogonality of an OSTBC provides
the fastest ML decoding (symbol-wise decoding) and the full diversity.
Unfortunately, this orthogonality is too strong so that the symbol rate
(symbols per channel use) of an OTSBC is upper bounded by $3/4$ for
more than $2$ transmit antennas and is conjectured to be upper
bounded by $(k+1)/(2k)$ for $n=2k-1$ or $2k$ transmit antennas \cite{haiquan1}.
This conjecture is true if no complex linear combinations
of two or more  information symbols is allowed in an OSTBC
\cite{liang1}. Systematic designs of OSTBCs achieving the conjectured
rate upper bound for an arbitrary number of transmit antennas
are presented in \cite{liang1, su, lufuxia}.
The designs in \cite{lufuxia} are inductive and have  
closed forms, while the other two in \cite{liang1, su}
are human-assisted or computer-assisted, and do not have closed forms.

\begin{figure}[t!]
\centering
\includegraphics[scale=0.55]{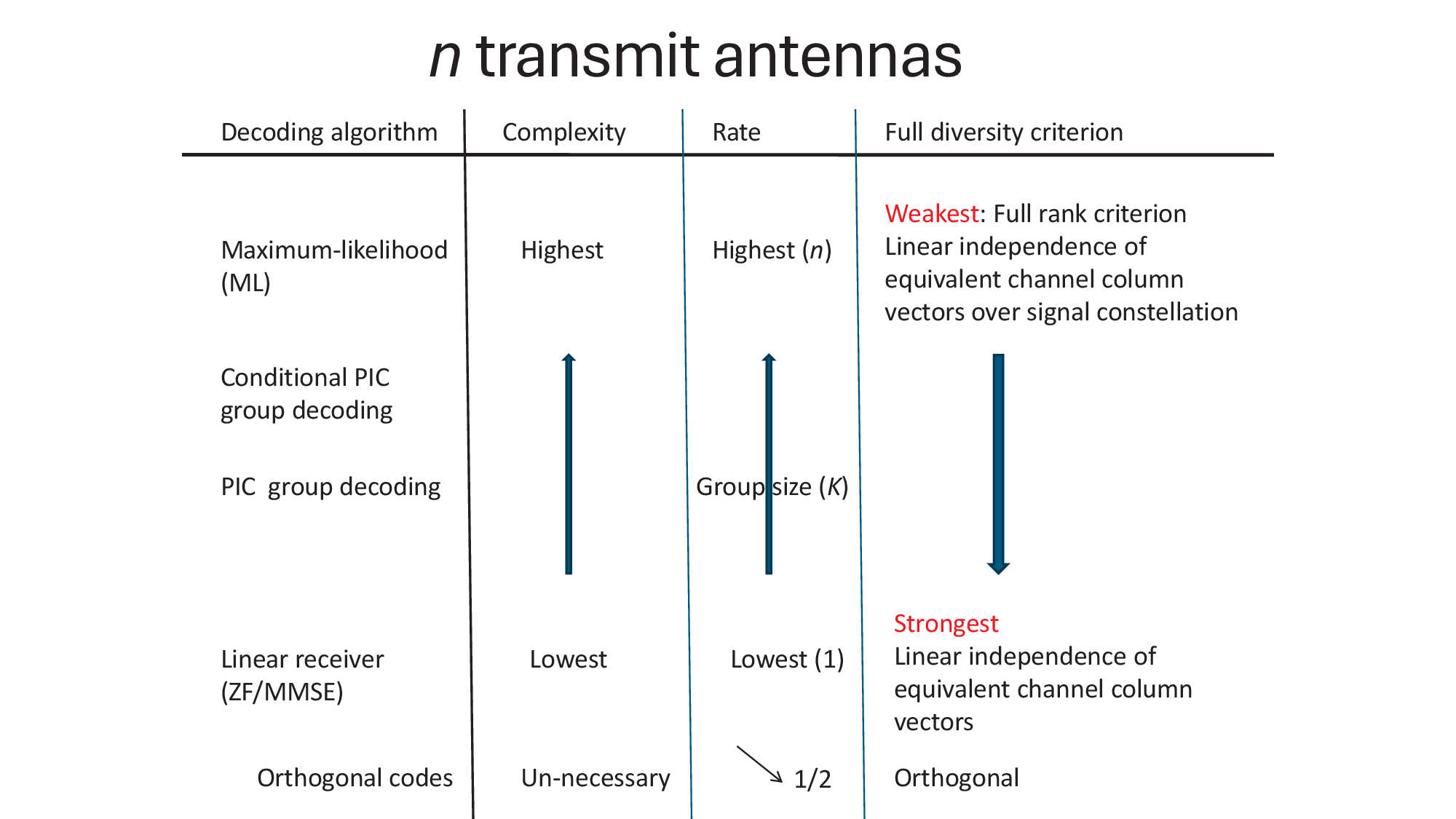}

\caption{Space-time coding tradeoff between decoding complexity,
  rate, and performance.}
\label{fig1}
\end{figure}

The above generally
mentioned TF/STF/ST coding with full diversity is based on ML 
demodulation/decoding.
Although OSTBC has fast ML decoding, their rates are low and approach
$1/2$ when the number of transmit antennas goes large.
However, for a low decoding complexity and having
full diversity in the meantime,
the orthogonality in an OSTBC is not necessary. 
Later we started to design ST codes
based on other low complexity demodulation algorithms, such as linear receiver
\cite{linear}, 
partial interference cancellation (PIC) group decoding \cite{pic1, pic2},
and conditional PIC group decoding \cite{cpic}. 
In summary, achieving a low complexity decoding and full diversity as well,
the sacrifice is the code rate. For example, the maximal code 
rate for linear receiver is $1$ \cite{linear}
and codes with rates approaching
$1$ and achieving full diversity with linear receiver are designed
in \cite{linear} as well. The maximal code rate for PIC group decoding is $K$
that is the group size \cite{pic1}.
One of the latest designs in this direction is \cite{shi2013}.
Fig. \ref{fig1} illustrates the tradeoff between decoding
complexity, code rate, and full diversity (i.e., performance). 
For more details, we would like to refer the reader to \cite{xia_talk}.

As a final note, ST coding has been combined with VOFDM in \cite{xia2, han, cheng} to achieve an improved performance over fading channels.

\section{Conclusion}
In this paper, we first showed that no modulation schemes can compensate
a non-trivial Doppler spread. The true
reason for all the claimed modulation schemes, such as
VOFDM/OTFS, are good for delay Doppler channels
is due to their block-wise/vector-wise demodulation in time or/and in
frequency, which
is good for general time-varying channels the same reason as BICM,
signal space diversity, and time-frequency coding. It is not because
they can specially  treat/compensate Doppler spread, and it is 
mis-leading in the community and  is clarified
in this paper.

We also summarized the main features of VOFDM and ST codes.
Since this paper is not about a comprehensive tutorial on ST codes,
a lot interesting works on ST/STF codes  over a decade ago
were not mentioned.

\end{document}